# TESTING THE LIMITS OF THE MAXWELL DISTRIBUTION OF VELOCITIES FOR ATOMS FLYING NEARLY PARALLEL TO THE WALLS OF A THIN CELL

**Petko Todorov[1] and Daniel Bloch[2]**

[1] *Institute of Electronics –Bulgarian Academy of Sciences, Sofia, Bulgaria*

[2] *Laboratoire de Physique des Lasers, UMR 7538 du CNRS,*
*Université Paris13-Sorbonne Paris Cité, 93430 Villetaneuse, France*

petkoatodorov@yahoo.com          daniel.bloch@univ-paris13.fr

*For a gas at thermal equilibrium, it is usually assumed that the velocity distribution follows an isotropic 3-dimensional Maxwell-Boltzmann (M-B) law. This assumption classically implies the assumption of a "cos θ" law for the flux of atoms leaving the surface, although such a law has no grounds in surface physics. In a variety of recently developed sub-Doppler laser spectroscopy techniques for gases one-dimensionally confined in a thin cell, the specific contribution of atoms moving nearly parallel to the boundary of the vapor container becomes essential. We report here on the implementation of an experiment to probe effectively the distribution of atomic velocities parallel to the windows for a thin (60 µm) Cs vapor cell. The principle of the set-up relies on a spatially separated pump-probe experiment, where the variations of the signal amplitude with the pump-probe separation provide the information on the velocity distribution. The experiment is performed in a sapphire cell on the Cs resonance line, which benefits of a long-lived hyperfine optical pumping. Presently, we can analyze specifically the density of atoms with slow normal velocities ~ 5-20 m/s, already corresponding to unusual grazing flight −at ~85°-88.5° from the normal to the surface− and no deviation from the M-B law is found within the limits of our elementary set-up. Finally we suggest tracks to explore more parallel velocities, when surface details -roughness or structure- and the atom-surface interaction should play a key role to restrict the applicability of a M-B-type distribution.*

**I. The Maxwell-Boltzmann distribution of atomic velocities in a dilute gas
and the signature of atoms moving close to a wall container.**

It is usually unquestioned, in the field of Atomic and Molecular Physics, that the atomic velocity distribution in a gas at thermal equilibrium is governed by a Maxwell-Boltzmann (M-B) distribution for kinetic energy, with a 3-dimensional isotropy. This is even the basis for proposals aiming to measure the Boltzmann constant by the Doppler broadening of a thermal gas, in view of a redefinition of the temperature scale [1]. However, in the vicinity of a wall, the isotropy is locally lost in a region, often called the Knudsen layer [2], where the atom-wall collisions modify the behavior of the gas. The microscopic description of an atom-surface collision or interaction is a rich field, which has been addressed experimentally for a long time. The possibility of a velocity distribution depending upon the vessel shape, has been sometimes considered, although the proofs for self-consistency seem delicate [3]. Despite the variety of behaviors found microscopically at the frontier of the gas, it remains usually assumed that the overall gas behavior obeys a M-B distribution (see *e.g.* [4]).

As recalled in detail in the review by Comsa and David [5], which includes a significant historical approach, the distribution of atomic velocities established by Maxwell for an isotropic volume of a gas with molecules undergoing numerous collisions (the basis for the ideal gas kinetics) has been founded upon questionable hypotheses concerning the effect of the surface. The Maxwell model assumes that a fraction *f* of the molecules sticks onto the wall, for further scattering or "desorption" according to the expected "gas law", while the remaining fraction (1-*f*) just undergoes an ideal (specular) reflection on the wall. Retrospectively, it is clear that specular atomic reflection cannot be considered as a general phenomenon for an unprepared surface and a high atomic density. Rather, quantum reflection on a wall potential is an effect well identified but difficult to observe, and applies to atoms cooled to extremely slow motion [6] (along the normal), while requiring perfectly polished surfaces [7]. For the desorption process, occurring on considerably longer time scale, arguments based upon conservation laws of the flux to and from the surface, combined with the assumed isotropic M-B distribution in the surrounding gas, lead to the $\cos\theta$ law (Knudsen law) for the flux of particles departing (desorbing) under an angle $\theta$, with $\theta$ the angle of the trajectory for desorption relatively to the normal (see notably [5], or the appendices of refs. [8] and [9]). Interestingly enough, the correct description of this input/output problem is susceptible to tackle the relevant description of such a general principle as the detailed balance [3,10], which is at the core of the thermodynamics theory [11]. Rectilinear and classical trajectories are implicitly assumed with $\cos\theta$ law, and this could also be questioned with respect to the microscopic details of a "surface potential", including spatial inhomogeneities. Also, when combining the dynamics of rarefied gas with a surface potential or with a particle-particle potential, the diffusion process, starting from the conditions of a gas at equilibrium, is susceptible to obey a Lévy statistics rather than a Maxwell law (see for a squeezed film of gas at low pressures [12] —experiments performed in view of technological applications, such as MEMS, gravitational waves detectors). Note however that diffusion processes may bring a too simple view as they are not bounded by a condition of return to the equilibrium, as usually assumed in gas kinetics.

Numerous microscopic studies of desorption [4,5,9, 13-15], have been performed by the community of physical chemistry or surface physics on well characterized surfaces under very good vacuum condition. Several studies show that, instead of the intuition of a dual behavior, with either specular reflection, either adsorption for an atom arriving onto a surface, one has to discriminate processes like direct inelastic or quasi elastic scattering, and trapping desorption mechanisms (see *e.g.*[16,17]). These processes obey very different time constants. The intrinsic corrugation of some surfaces has also to be considered (*e.g.* corrugation by aromatic molecules [18]), and may even bring more consequences for a gas enclosed by micro- or nanostructured surfaces. In a variety of systems, one finds that atoms or molecules are susceptible to leave the surface with superthermal mean velocities [18], or in other situations with subthermal velocities [8]. These mean velocities are mostly defined for an observation selecting only one velocity component. Hence, it is now clearly viewed that desorption processes are far from always obeying the cos $\theta$ law [5,15]. It is notably frequent to find a sharper peak around the normal [14], with a distribution traditionnally approached by a $\cos^n \theta$ law, with n > 1. These experimental studies of desorption have been unable to study nearly "grazing" desorption or scattering, the technical limitations appearing around $\theta \sim 70°$, or sometimes 80°.

The development of laser spectroscopy techniques, easily allowing selection of atomic or molecular velocities and quantum levels [9], should have helped to study in depth the cos $\theta$ assumption of the gas kinetics theory, which is clearly oversimplified with respect to the current development in surface sciences. To the best of our knowledge, if one restricts to the case of an atomic gas at equilibrium, hence avoiding to populate a variety of rotation/vibration levels as occurs with molecules [4, 9,14-15], the problem was addressed experimentally only by Grischkowsky [20]. In this 1980 work, the extensive bibliography includes experimental references from the early times of the 20[th] century, whose validity may be retrospectively questionable because of concerns about the vacuum quality (possibly limiting free flight trajectory) and about the nature of the surface. The Grischkowsky experiments are limited to an agreement with the predictions for the cos $\theta$ law, with sensitive discrepancies if an alternate $\cos^2\theta$ law is considered. The experiment was performed on a low-density thermal vapor of sodium contained in a cylindrical glass cell traversed on-axis by a resonant laser beam. The information on the velocity distribution of atoms leaving the walls of the cylindrical cell is brought by the transmission spectrum, which integrates the Doppler shift with respect to the velocity distribution, while the long free path of the Na atom in the low-density vapor enables to observe atoms flying from the wall. The cell dimensions -whose anisotropy remains limited, with a 50 cm length and 2.5 cm in diameter-, and the long distance between the walls and the optical probe, are however inappropriate to explore the specific response of atoms moving nearly parallel to the wall. For a similar system of glass and Na atom, a specific observation of desorbing atoms through 2-photon spectroscopy [21] has also claimed to show agreement with the $\cos\theta$ law for the desorbing flux. The proof is actually based upon a better agreement between the experiments and the predictions for the $\cos\theta$ law, than if one models an isotropic flux, corresponding to an unexpected $(\cos\theta)^l$ desorption law with $l = 0$ (*i.e.* assuming an already 3-dimension thermalized gas at very small distances from the surface). Here again, the experiment is unable to explore the distribution of atoms desorbing under a near grazing incidence (*i.e.* large values of $\theta$). Note also that in the two works mentioned above, the surface was simple polished glass, and the experiments are far from providing "universal" evidences of the effectiveness of a "$\cos\theta$" law.

A blooming of sub-Doppler resolution spectroscopy techniques in a thin film of dilute vapor at thermal equilibrium has been noted in the last decades, including selective reflection spectroscopy at an interface [22,26], and confined vapor spectroscopy in micro- or nano-cells [22,27-34]. The Doppler broadening, associated to the distribution of random atomic velocities, can be eliminated for well-chosen signals –note by the way that the early observation of a sub-Doppler contribution in selective reflection [35], before the laser era, was originally attributed to the contribution of specular atomic collisions onto the surface, so that the first modern theory [36] of selective reflection spectroscopy also includes the specific contributions of atoms having undergone specular reflection. In these techniques, usually requiring a single laser beam, the sub-Doppler signals appear for an irradiation under normal incidence. Indeed, only atoms with a "slow" motion, as counted along the normal to the "wall" (or vapor container), are unaffected by the deleterious effect of a transient response to the exciting light (assuming that the atom-light interaction is interrupted by a wall collision, or starts from scratch for departing atoms [37]). A major condition to make sub-Doppler signals observable is that the gas is dilute enough for atomic trajectories to be determined by collisions with the wall, rather than by atom-atom collisions as usual for an isotropic three-dimensional gas. This allows the spectroscopic response of "slow" atoms to be enhanced relatively to the Doppler-broadened response. This specific contribution of the atoms travelling near parallel to the window can even be singled out in various cases through an appropriate processing of the lineshape, notably. a frequency derivation of the lineshape [23,24,29] by applying a Frequency Modulation (FM) followed by the coherent demodulation of the signal. In several cases, this processing is required to single out the zero (normal) velocity group $v^{\perp}=0$, whose relative width, with respect to the thermal velocity $u_{\text{th}}$, is the ratio of the optical width $\gamma$ to the characteristic Doppler width $\Gamma = k u_{\text{th}}$ (with $k$ the wave vector of the irradiation). In a two-laser nonlinear selective reflection spectroscopy experiment [25], where an arbitrary selection of normal velocities can be operated, the analysis of the signal amplitude has seemsed to imply that the number of arriving atoms at grazing incidence ("slow", relatively to the normal) is smaller than the number of departing ones at the same large incidences. This was interpreted as an acceleration of arriving atoms, *vs.* trapping for the departing ones, as due to the surface interaction. Velocity selection can also be operated through transient optical pumping technique (pumping towards a third-level for which the irradiation is non resonant), and "slow" atoms are thus defined as those travelling less than the observation zone during the time ($\tau_{pump}$) required for the optical pumping to take place. This time can be quite long time at low power [27,28] apparently implying a sharp velocity selection. The effective weighting of slow ($< u_{th} \ \gamma/\Gamma$) or very slow ($< [k\tau_{pump}]^{-1}$) atoms, analyzed in detail in [30], remains actually dependent on the optical linewidth of the transition.

Experimentally, these techniques were implemented only for alkali-vapor lines — on resonance lines, or sometimes on weak and more excited transitions [22,25,26,31,33]—, but always in a regime where collisions impose a notable broadening to the optical width $\gamma$. Hence, the gain in resolution, relatively to Doppler-broadened spectroscopy, has always remained well below 2 order of magnitudes, and the selected atomic (normal) velocities were at best on the order of 5-10 m/s —to be compared with a thermal velocity ~200 m/s for Cs. The possibility of benefiting of a more stringent velocity selection through narrow transitions, in view of references of a metrological interest, has been considered in a theoretical proposal (on the intercombination line of Ca) [38], or briefly attempted experimentally (on the intercombination line of Ba —see page 128 of ref. [22]). The velocity selection could hence correspond to atoms flying, in a one dimension direction, as slowly as $v^{\perp} \leq$ cm/s or less in the lab frame (*i.e.* vapor cell frame), an extremely slow velocity attained only at the laser cooling era, easily showing sensitivity to recoil effects –in the presence of an interacting light. The difficulty for using atoms at such slow velocities is that one here assumes vicinity with a macroscopic container. Such a situation strongly differs from the commonly sharp velocity selection operated in non linear spectroscopy (*e.g.* saturated absorption spectroscopy) when an arbitrary narrow selection can be operated in a frame defined by an optical frequency difference. In other words, if describing the gas in the lab frame or in a moving frame can be chosen with no consequence for an isolated gas far from the container, the lab frame has a profound meaning when atom-surface interaction or collision is considered. A natural consequence of an assumed M-B distribution of atomic velocities is that the distribution of "slow" atoms (for the normal component $v^{\perp}$, *i.e.* atoms moving parallel to the wall) should be flat around null velocity. However, when pretending to use those atoms moving nearly parallel to a cell window, intrinsic difficulties arise: the long-range atom-surface (van der Waals type) attraction may curve the atomic trajectories, and this should particularly affects atoms moving parallel to the wall [22, 33]; also, the local roughness of the window, which often largely exceeds an atomic step, may prevent very parallel motion to the window plane.

In the principle, the development of experiments involving linear spectroscopy of a vapor close to an interface [22-24, 29-31], even when limited to relatively broad sub-Doppler transitions, should have indicated to which extent the velocity distribution is flat for velocities around the selected velocities, currently 5-10 m/s. This should require analyzing the signal amplitude as a function of the broadening (*e.g.* collisional broadening), especially if the sub-Doppler transition width can be related to the velocity selection. Single beam nonlinear spectroscopy in a thin cell has also been considered [28,30] for an effective quantitative investigation of the presence of slow atoms. For various reasons, this amplitude study, although highly significant in the context of *linear* spectroscopy, has never been performed systematically [39]. In a very brief extension [31] of the pioneering work on thin cell spectroscopy, a spatially separated pump-probe technique had been implemented, and the observed optical response was attributed to Cs atoms with a normal velocity ≤ 2 m/s (to be compared with $u_{th} \sim 200$ m/s —note that, for this 2m/s claim, the absence of residual beam overlap was not investigated in-depth). A critical constraint on the cell fabrication, not achieved at that time, is the cell parallelism over the whole pump-probe separation. Further works with spatially separated pump-beams [42,43] had been concerned with linewidth narrowing, but had never addressed the quantitative measuring of slow atoms.

Here, we report on the feasibility of measurements with a dedicated set-up, based upon an experiment with a thin (60μm) Cs vapor cell. A ring of pump light, whose diameter is adjustable, induces a long-lived signature in the atomic medium, which is interrogated on a central probe spot to observe atoms having moved nearly parallel to the window. After describing (section II) the principle of the set-up, we report on the experimental results (section III) addressing incidences as high as 85-88.5°, and finally discuss (section IV) the tracks to overcome the limits of the present measurements, in order to address the atomic distribution at even more grazing incidences.

## II. Experimental principle and set-up description

The set-up, whose build-up has been reported in [42], consists in a saturated absorption experiment with spatially separated beams performed under normal incidence on a thin vapor cell (thickness: L) whose walls are parallel (see fig.1). For low vapor density, atomic trajectories are essentially wall-to-wall trajectories and atom-atom collisions should be negligible. The principle of the experiment is that the probe beam response originates only in atoms having travelled (nearly) parallel to the wall for more than the long distance R, from the pump region to the probe spot, while their normal motion has not exceeded the small thickness L of the narrow cell. The pump-induced variation of the probe transmission, measured as a function of the pump-probe distance R, provides the information on the effective distribution of atomic velocities. In other words, the angle L/R, which can be made quite small in a microcell (typically $L \leq 100\,\mu m$) for a macroscopic pump-probe separation ($R \gg L$ *e.g.* $R > 1$ mm), defines the accuracy with which parallel atomic trajectories are investigated. Typically, for an atom whose velocity modulus is just the thermal velocity $u_{th}$ (instead of a distribution around $u_{th}$), the set-up selects the response of atoms whose velocity, normally to the wall, satisfies $v^{\perp} \leq u_{th}\,L/R$. A more precise evaluation ([42] and fig.2) softens this rough evaluation. To "count" the distribution of those "slow" atoms, one needs to perform a high-efficiency pumping in the pump region (fig.1), lasting sufficiently for an atom to reach the probe spot (optical pumping in the ground state for an alkali metal vapor is very convenient for this purpose). Also, pumped atoms which are to be detected in the probe spot should not be destroyed, nor reinjected, through collision processes (atom-atom, or atom-wall as well).

Our experiments are performed on Cs vapor, whose resonance lines doublet —we explore here the $D_2$ line $6S_{1/2} \rightarrow 6P_{3/2}$ line at 852 nm— are known to be strong, inducing an efficient depopulation of the ground state. Despite the short lifetime (~30 ns) of the optically excited state 6P, a saturating beam easily provides an efficient population transfer (~100% transfer after a few cycles of optical pumping) to the "other" hyperfine ground state F= 3 or F =4 (whose 9.192 GHz separation is the basis to define the Cs clock), transparent to the selective resonant excitation. The relaxation of this optical hyperfine pumping is usually extremely long in the vapor itself (it occurs essentially through wall collisions, in the absence of a strong applied magnetic field). Usually, a single wall collision on a glass tube is not sufficient to fully thermalize the hyperfine components of the ground state [43].

Specific coatings, such as paraffine or more recently alkene material -see [44]-, can now prevent the relaxation (for up to even $10^5$ collisions) : they should definitely not be used for the present experiments. Hence, if it is easy to realize an efficient "mark" (through the optical pumping) of the atoms having left the pump region to reach the probe spot, one cannot ignore that "pump-marked" atoms may reach the probe spot after a complex trajectory, with their velocity redistributed by atom-atom or atom-surface collisions. Hopefully, the intrinsic anisotropy of the thin cell geometry makes diffusive propagation to the probe region rather inefficient. For atomic trajectories bouncing onto the (glass) walls of the cell, a rather long dwell time on the surface (~ several μs [43]) has often to be accounted for, which may impose an additional long delay between the pumping interaction and the directed diffusion towards the probe spot. Again, the historical and somehow naive discrimination between specular reflection and adsorption tends now to be amended, with respect to experimental evidences, by a discrimination between a (long) trapping of slow atoms and a diffusive bouncing of faster atoms [17].

To optimize the efficiency, we have chosen a ring-shaped pump, whose optical intensity must be sufficient to perform a strong hyperfine pumping —towards the long-lived other hyperfine sublevel of the ground state— while the probe is sent at the centre of the ring. Such a geometry is inverted with respect to the initial and preliminary demonstration [31] of spatially separated pump-probe experiment in the context of a thin cell, and is analogous to the one proposed in [40]. The advantages of this geometry is that all sufficiently "slow" atoms (for the normal component of the velocity) entering through the probe spot should have undergone pumping in the pump ring region. The constraint is that the ring-shaped pump should be intense enough to provide a full pumping, or at least a constant pumping rate, for the atoms leaving the inner part of the ring, even when the inner diameter R of the ring is increased to explore smaller and smaller velocities. In such a scheme, and provided that the velocity distribution for "slow" atoms (normally to the wall) remains flat, the pump-induced probe signal should simply evolve as $R^{-1}$. Conversely, if very slow atoms $v^{\perp} \sim 0$ have disappeared (because of surface roughness, curvature of trajectories, ...), the signal should drop down faster than $R^{-1}$. It could even reach directly zero for a typical pump-probe distance $R \geq L u_{th}/v^{\perp}_{min}$, for $v^{\perp}_{min}$ the minimal normal velocity, if it appears that parallel velocities are truly prohibited, and if one assumes for the atomic kinetic energy distribution a quickly decaying tail for velocities exceeding $u_{th}$, as for a M-B distribution. More detailed predictions on this number of probed atoms as a function of the pump-probe distance were calculated with late Prof Saltiel, in the case of a flat hole in the distribution of slow (normal) atoms, and reported in [42]. Figure 3, extrapolated from [42], shows how the absence of slow atoms would appear in the experimental scheme that we consider. The independent analysis provided in [40], for the geometry that we use presently, was restricted to the limiting assumption of an ideal M-B distribution, and oriented to lineshape peculiarities with respect to the finite size of the pump beam ring.

To count the number of atoms which have travelled from the pump to the probe, or at least to count them *relatively*, *i.e.* the same way whatever is the pump-probe separation, the intensities of the pump and the probe are an issue. Ideally, the pump ring should be saturating enough so that all atoms moving from the inner diameter R to reach the probe spot are pump-marked. Care must be taken when increasing the pump diameter not to decrease the highly efficient pumping for atoms moving from the pump region to the probe spot. For the probe beam, whose absorption is measured, and which remains unchanged when varying the separation R, it is preferable to use a weak non-saturating intensity, as saturation decreases the relative absorption. Nevertheless, the probe intensity must be sufficient to avoid the intensity-dependent transient regime of light-atom interaction, which corresponds to the delayed appearance of absorption. This regime, although clearly observable [29,30], essentially applies for vapor cells of a very small thickness, on the order of the optical wavelength $\lambda$ itself.

In the experimental set-up, the thin vapor cell has an internal thickness L = 60 $\mu$m and ~2 x 3 cm in size, with a Cs drop deposited in a bottom reservoir (see fig.4). It is used instead of the initial 19 $\mu$m-thick cell, mentioned at the early times of the set-up [42]. The windows are made of a standard high quality polished sapphire (estimated roughness $\leq$ 5 nm). The local variations of the thickness were tested interferometrically, and do not exceed 1 $\mu$m over the whole cell diameter. The deviation to parallelism is less than 100 nm over 1 mm (*i.e.* parallelism as good as $10^{-4}$ rad). The cell (and the Cs reservoir) is heated up to a temperature ~ 40 °C in order to increase the Cs atomic density ($\leq 2.10^{11}$at/cm$^3$). A moderate temperature change should not affect the thickness and the parallelism.

The laser beams are provided by a single diode laser (DFB type, short-term linewidth ~4 MHz), temperature-controlled, whose frequency is scanned around the Cs $D_2$ line (852 nm) by the driving current. The laser beam is divided in two counterpropagating beams (fig. 5): the probe beam goes directly through the thin cell —under normal or near-normal incidence—, while the pump beam is first directed onto an optical system, which converts the approximately Gaussian beam into a conical beam, emerging under a half-angle 11°. The optical system consists of a sharp focusing of the pump onto a circular grove grating —period 4 $\mu$m, from Coherent Canada Inc. — which produces a diverging cone of first-order diffraction, whose divergence is reduced by an axicon. The conversion efficiency in the first order ring is about 50%. A small blocking disk is introduced on the path of the pump beam after the axicon, and in front of the cell, to ensure that the zero[th] order diffraction, although weak, is eliminated. Aside from a better elimination of spurious light originating directly from the pump, this has been the practical reason to choose a nearly counter-propagating geometry for pump and probe [45]. Note that the holder of this disk blocks a small fraction of the incident pump ring before the cell is reached. After emerging from the axicon lens, the pump beam is still divergent so that it is focused by a converging lens. The thin cell is situated after the focal point of the pump in order to avoid stray pump reflections towards the probe region. Changing the pump-probe distance R is simply achieved by moving the cell parallel to the probe or the (aligned) pump-cone axis. Note that despite the oblique incidence of the pump, the cell thickness is small enough for the distance between the axial probe and the inner diameter of the pump ring, to be constant over the cell.

The probe spot was chosen to have a diameter ~ 0.5 mm, as obtained with a slight focusing. There is indeed a trade-off between the resolution limit to the pump-probe separation distance R, and the signal amplitude (enlarged when increasing the size of the probe spot). The pump inner ring is conveniently measured through its angular divergence after the focal point of the converging lens, and by varying the distance between the focal point and the vapor zone in the cell (corrections due to the optical length of the cell windows are negligible).

Conceptually, the experiment simply relies on the measurement of a pump-probe signal. In its principle, it does not even require the analysis of a spectrum: one only needs the pump beam to be sufficiently resonant to effectively modify the atomic state, and the probe should be tuned to an atomic resonance (the same as the one for the pump, or a different one) solely to count the pumped atoms. In our experiment, we use the same laser for the pump and the probe beam, and the laser frequency is usually scanned across one of the hyperfine manifolds of the $D_2$ transition (F=3→F'={2,3,4} or F=4→F'={3,4,5}), with respectively 150 MHz and 200 MHz hyperfine structure, or a 200 MHz and 250 MHz hyperfine structure [46]. Under normal incidence (as chosen for the probe), only a Doppler-free signal is expected to be recorded. Because the pump is also rather close from normal incidence, the pump irradiation has a nearly optimal efficiency for the atoms of interest — whose motion is parallel to the cell windows. As we shall see in the next section, this frequency selectivity allows to eliminate the effect of a residual pump-probe overlap; it tends in addition to warranty that the probe detects only slow atoms when crossing the probe, and not "marked" atoms whose velocity would have been modified by collisions. Also, because the signal is expected to be weak, an auxiliary set-up is implemented, to test the absorption and the sub-Doppler saturated absorption on a macroscopic Cs cell at room temperature.

At last, the set-up requires an on-off modulation of the pump beam, and a correlated detection on the probe, usually performed with a lock-in detection. The modulation was initially performed with a (slow) chopper, and has been replaced by an acousto-optic modulator (AOM), for a modulation at 1.3 kHz. In this last case, it is needed to use the zero[th] order diffracted beam for the modulated pump, in order not to shift the pump frequency by the acoustic frequency (an amount which usually corresponds to the Doppler shift of rather fast atoms). Care should be taken for the pump modulation to reach a quasi-total extinction (the residual value is ~8 %) on the zero[th] order beam : indeed, for a strongly saturating pump beam, a small fraction of the incident pump beam may be sufficient for an efficient pumping, leading to a notable reduction of the modulation appearing on the probe beam.

The experiments were most often conducted with a probe beam intensity ~ 1 µW, ensuring that it is largely non saturating (~ 16 µW/mm² is the saturation intensity for the $D_2$ line, see [46]) while the pump beam was typically about 1.5 - 2 mW, distributed on a ring whose thickness is a fraction (~ 0.2) of the diameter (fig. 6). The ring internal diameter was varied but to not more than 4 mm (*i.e.* 2 mm in radius), so that at maximal available power, the pump was sufficiently saturating even for the larger inner diameter, as shown by fig. 7 (the figure shows the results only for the F=4→F'=4 transition, and similar results are found for all hyperfine components).

## III Experimental results

First, when there is an overlap between the pump and the probe beam, at least partially, the situation is essentially similar to a usual saturated absorption experiment. Such a situation of course happens by positioning the cell at the focal point of the pump. It can also occur through residual reflections if the pump beam converges onto the cell, or may simply occur with a partial overlap if the pump-probe separation is not sufficient. The demodulated signal observed on the probe beam is simply a saturated absorption signal, characterized on the $D_2$ line by the appearance of main resonance lines —corresponding to the signature at null (normal) velocity— and of cross-over resonances associated to atoms whose normal velocity allows the respective Doppler shifts for the pump and the probe to compensate: the velocity associated to the crossover resonance is $v^\perp$ ~75 m/s for the F'=2-F'=3 crossover, $v^\perp$ ~100 m/s for the F'=3-F'=4 crossover, $v^\perp$ ~125 m/s for the F'=4-F'=5 crossover, or respectively $v^\perp$ ~175 m/s for the F'=2-F'=4 crossover and $v^\perp$ ~225 m/s for the F'=3-F'=5 crossover which are hardly isolated from the main respective resonances F=3→F'=3 and F=4→F'=4. Although the cell is a "thin cell" (the vapor is dilute enough for wall-to-wall trajectories, and the irradiating beam diameter exceeds the cell thickness), the crossover resonances are not that much attenuated by the lower efficiency of fast (normal velocity) atoms [30]. The width of the sub-Doppler peaks, should be at minimum 5 MHz (linewidth of the transition). It is actually broadened by the accumulated fluctuations of the (free-running) laser frequency during a scan, and because of various collisions processes (impurities, velocity changing collisions). In addition, in the case of a pump-probe overlap, one can expect a residual Doppler broadening, coming from the non-normal incidence of the conical pump (due to the small thickness, the pump ring can overlap with the probe beam over the whole cell thickness in spite of the nonparallel geometry). As usual, the amplitude of the "cycling" transitions (namely $F = 3 \rightarrow F' = 2$ and $F = 4 \rightarrow F' = 5$), which are normally unable to bring the atom into the transparent hyperfine component of the ground state level (respectively $F = 4$, and $F = 3$) is weaker than for other transitions, because for these cycling components, the memory of the pumping remains short. This is in spite of a higher transition amplitude for $F = 4 \rightarrow F' = 5$ (in a ratio 44 to 21 for $F = 4 \rightarrow F' = 4$ and 44 to 7 for $F = 4 \rightarrow F' = 3$ —see [45]). In addition, the resolved sub-Doppler peaks appear on a relatively broad background, originating in a redistribution —up to thermalization— of the velocity by collisions, yielding an observable signature of pumped atoms on the probe beam. This background is approximately as broad as the Doppler width, as if it corresponds to full velocity redistribution. Hence it can originate as well in atom-atom collisions —possibly conserving partly the atomic velocity when considering only a single collision process [47]— or in atom-wall collisions, commonly associated to a full redistribution of atomic velocities. In all cases, the narrow peaks correspond to an increased probe transmission (or reduced absorption), as the pump beam has selectively transferred absorbing atoms into the transparent hyperfine sublevel.

Separation of the pump and probe beams is achieved by moving the thin Cs cell from the region of optimal overlap, characterized by a maximal signal. This ensures that the other parameters of the set-up (power of the beams, cell temperature,...) are kept unchanged —provided that the thin cell is transversally homogeneous. A sensitive variation of the pump ring diameter is obtained by the cell longitudinal position. For an improved accuracy on the pump-probe separation, we take consideration of the finite size of the probe, for R to be the true pump-probe separation. The spectra of fig. 8 shows how the signal amplitude decreases when the separation is increased. The absence of the crossover resonances is a clear signature of the absence of a residual pump-probe overlap. Meanwhile, a broad background remains observable, indicating that some pumped atoms, of any velocity, reach the probe spot after a complex path, including a number of wall collisions, where they have not undergone a fully relaxing wall-collision, hence keeping some memory of the pumping step. Note that the cycling transitions (F=4→F'=5, and F=3→F'=2) yield smaller signals, but not null signals as one would expect for a genuine two-level system with a lifetime much shorter than the transit from pump to probe. This is because the pump saturation, along with the residual Doppler broadening on the pump laser, allows hyperfine ground state transfer through the detuned excitation to a third level (*e.g.* a laser tuned to the F=4→F'=5 transition induces the F=4→F'=4 transition, although with a lesser efficiency).

To analyze the data, we measure the amplitude of the main narrow lines, after subtracting the contribution of the residual broad background, as appearing in fig.8. This solves a principle difficulty: as mentioned in section II, the signature of atoms in free flight from pump-to-probe may be mixed-up with the one of atoms travelling from pump-to-probe while undergoing collision processes. Here the spectroscopic signature of those collisions, differing from the narrower signature of free-flight atoms, allows to remove this unwanted contribution. After such a data processing, we have collected the amplitude of all the six hyperfine components as a function of the pump-probe spacing. Figure 9, which converts the spectra of fig.8 into peak values of the amplitude after subtraction of the background, confirms for all hyperfine components the rapid drop of the signal when increasing R, already apparent on fig.8. When normalizing the fig.9 data as in fig.3, in order to compensate for the decrease in amplitude naturally associated with the sharper velocity selection when increasing the pump-probe separation R/L, one finds (fig.10) a nearly flat behavior for all hyperfine components. The individual error bars in fig 9 are large because of the noise level affecting peak amplitudes, and because of a notable uncertainty when subtracting an estimated background —in the absence of a lineshape analysis, or of a model allowing a precise evaluation of the background. For fig. 10, the normalization is applied with the R/L factor and ignores the initial ratio between each hyperfine component, so that the unity results statistically from an averaging for various separations R. This allows reducing the error bars, relatively to fig .9, despite the residual uncertainty in the resolution of the spatial separation (*i.e* in the R/L factor). Around these normalized values, a linear fitting shows a quite flat distribution as a function of the pump-probe separation, with remains well inside the error bars, and with a slope whose signe varies from one hyperfine component to another one.

From our study on a 60 µm thick cell and a pump-probe separation explored in the ~0.5-2 mm range, we can conclude that the velocity distribution is compatible with a flat distribution for normal velocities between 20m/s to 5-10m/s. Although such an information, limited by large error bars, may seem solely to confirms a common hypothesis in atomic spectroscopy, it is worth emphasizing that this had never been investigated systematically before. With respect to a thermal velocity ~200 m/s for Cs atoms, our investigation demonstrates consistently the feasibility of investigating trajectories at a grazing incidence 85°-88.5°, not attained by common methods for beams of desorbing particles. The major limit in this apparent agreement with the M-B distribution is that the investigated velocities $v^{\perp}$ are not extremely slow. Until now, our investigation of slower velocities was made difficult because the signal tends to drop down below the noise when the pump-probe separation exceeds ~2 mm. A thinner cell may seem to allow a more ambitious analysis of very grazing velocities (*i.e.* higher R/L values). Actually, our very preliminary investigations [42] were performed on a thinner cell (L = 19 µm), but they had suffered from a residual pump-probe overlap. In addition, the use of a thinner cell to allow a more stringent velocity selection for a given R separation requires (for sensitivity reasons) increasing Cs temperature —and density— to compensate for the smaller signal, at the expense of increasing the contribution of Cs-Cs collisions.

In the present set of experiments, our spectroscopy set-up remains rather elementary, despite the original implementation of a ring-shaped pump beam. It was indeed intended to demonstrate the feasibility of the analysis of slow velocities. In the next and final section, we show that various improvements could be realized, which should allow to decrease the range of atomic normal velocities ("slow", or "parallel" atoms) to be investigated for specific systems of interest.

## IV. Towards improved constraints on the density of slower velocities

The main limitation that we have encountered is the drop of the signal amplitude when the pump-probe separation is increased. Indeed, assuming a standard M-B distribution (*i.e.* flat for small velocities), the amplitude is inversely proportional to this separation (see figs. 3, and also figs 9-10), and the drop is even more severe if atoms with very parallel velocities cannot travel over long distances. We discuss here various improvements which may allow to check quantitatively the presence of slower atoms, notably the limits in the signal-to-noise ratio and the advantages of a dual laser set-up, along with the effects of the physical nature of the surface

Until now, the main limit to sensitivity is in the signal level itself, rather than an intrinsic limit in the signal-to-noise ratio. The noise level on the probe beam mostly originates in the amplitude noise of the laser source. It is sensitive to some technical noise (drive current stability, mechanical stability of the spatially filtered beam, at the frequency of the pump modulation), but it could be reasonably lowered down to the shot-noise limit for an optimal laser. Hence, improvements in the stability (laser source, mechanical set-up), and data accumulation should be a natural track for a better sensitivity. In addition, increasing the probe power would allow to collect more photons. To avoid the probe to be saturating, an enlarged probe diameter can be used, and this would only affect marginally the spatial resolution in the R distance, which is a non critical issue as long as the pump-probe separation largely exceeds the probe diameter. For the present set-up, all spectra were recorded with a free-running laser under a rather fast scan (120-300 s for a ~500 MHz scan, as imposed by the frequency drifts of the laser), so that less than few seconds were spent on each of the significant sub-Doppler peaks. In its principle, the measurement demands mostly to record the amplitude at a single frequency, allowing considerably longer recording time, provided the laser frequency is well stabilized. Even if the recording of a full lineshape appears needed to substract a Doppler-broadened collision background —see section III—, only a limited number of well controlled frequencies should be truly useful, and only the recording time at a given frequencies may be accumulated for times much longer than in our present study. This should make currently attainable an increase of sensitivity by an order of magnitude (*i.e.* velocity ~ 1m/s), allowing to increase correspondingly the R/L value range (assuming a flat distribution of slow atoms).

Replacing the single DFB laser set-up with the higher flexibility of a dual laser set-up (*i.e.* a powerful pump laser and an independent tuneable narrow frequency-linewidth probe laser) would bring notable advantages to the set-up. Aside from technical advantages—easier elimination of spurious interferences on the probe beam, possibility of using a strong or amplified pump beam—, the independent tuneability of the probe frequency should bring stronger constraints on the detected pump-probe signal. In particular, for a given pumping (at a fixed frequency and intensity) of the hyperfine ground state, the linear probe absorption should simply follow the relative amplitude of the hyperfine components (of the excited state), instead of exhibiting a nonlinear dependence on the pumping efficiency as in the present set-up. Also, with a truly narrow-frequency probe laser, one should observe a spectrum as narrow as the pure Doppler-free transition (*i.e.* 5 MHz for Cs), instead of the rather wide spectrum (FWHM 25-40 MHz) that we obtain presently, because of instabilities and drift in the frequency scanning process. Observing a pure Doppler-free transition without broadening would allow to rule out the possible influence of residual atomic collisions by impurities (Cs-Cs collisions are usually estimated to be negligible at our low temperatures). Also, based upon fig. 2 or on an elaborate calculation of the distribution of atoms reaching the probe region, the probe transmission lineshape is expected to be the convolution of the Lorentzian resonant absorption by the slight Doppler-shifts associated to the peaked distribution of selected velocities. Hence, for moderate pump-probe separations, a detailed analysis (deconvolution) of this lineshape, with its tiny residual Doppler broadening, should yield valuable information for the distribution of nearly grazing velocities. Shifting the pump laser frequency may also allow to selectively pump atoms colliding the wall soon after pumping, opening the way to evaluate the amount of pumped atoms reaching the probe spot after at least one wall collision process. This may provide a more sensible evaluation of collision effects than our removal of the broad collision background (section III), and consistency of the collision signal with the pump-probe separation could be checked. In this case, a smaller cone angle for the pump (after the circular grating/axicon sytem) would be worth considering to allow for a more precise selectivity in the initially pumped velocity.

At last, the choice of the surface has also to be considered with care. The technique that we have developed intrinsically allows an *in situ* measurement once the adequate shape cell is chosen and over the pump-probe distance separation. The nature of the window (sapphire here) is susceptible to determine the availability of atoms with very parallel trajectories: this is because of the long-range surface attraction itself or in consideration of possible specific trapping/adsorption mechanisms. The microscopic roughness of the surface, as well as the quality of the macroscopic optical polishing —in addition to the parallelism of the microcell windows, required for our measurements—are susceptible to have a dramatic influence on the distribution of atoms leaving the surface with $v^\perp \sim 0$ [48]. Super-polished annealed sapphire has already shown improvements in the reproducibility of atom-surface measurements [26], and other types of superpolished windows (quartz, glass, ..) are now also available. In addition, future thin cells to investigate the presence of slow atoms should also be free of residual impurities as much as possible, a task which has its own difficulties because of the high impedance of the specially designed cell.

**CONCLUSION**

By observing a spectroscopic signal originating in atoms having travelled up to 2 mm of a free flight in a cell of a microscopic thickness (60 µm), we show that information on the effective velocity distribution for trajectories nearly parallel to a surface can be reached. Until now, thin cell spectroscopy had relied on the contribution of rather slow atoms, but constraints on their distribution had never been looked for. With our demonstration set-up, which does not allow long integration times, the current experimental uncertainties do not show a significant deviation to the M-B expectation for the 5-20 m/s range of normal velocities (*i.e.* incidence angle ~85°-88.5° from the normal). This is in spite of the absence of firm grounds to build-up the M-B distribution in the close vicinity of a surface, and can be already considered as an unprecedented experimental result. For a further generation of similar experiments, one can reasonably expect that the velocity distribution could be addressed down to ~1 m/s, opening the possibility that the results depend on the properties of the surface: the chemical nature of the window (*e.g.* sapphire *vs.* glass, or also implementation of a non opaque metal film in the pump region), the quality of its polishing or the possibility of an imprinted nano-structuration, or its temperature as well, are indeed susceptible to alter the atomic desorption and the resulting distribution of atomic velocities. Even if until now we had considered only crude deviations from the M-B law —eliminating all atoms slower than a threshold—, it is clear from fig. 3 that any notable deviation to the flat M-B law (for slow atoms) should bring a notable change in the dependence of the signal amplitude as a function of the R/L parameter, notably when R/L can reach 100-200 values.

Extensions to different species, notably molecules, could be of interest, allowing to combine the rovibrational selectivity intrinsic to spectroscopy with the directional selectivity (for sharply grazing incidences) offered by the spatially separated pump-probe technique. For most molecular gases, the temperature of the surface (or of the whole cell body) can be changed without affecting the molecular gas density —or pressure—, so that simple processes can be sufficient to build-up the thin gas cell. A study of this influence of the surface temperature, with respect to the slow atoms of a given rovibrational level would be of interest, noting that it is frequent that molecule-surface interaction does not equally "thermalize" rotation, vibration, and translation [5, 50, 51]

A general difficulty with the principle of our set-up may be found in the competition between the observation of atoms with a very infrequent trajectory, and the possibility that atom-atom collisions, occurring somewhere in the "middle" of the cell, release atoms with "parallel" trajectories, which keep memory of their pumping in spite of having collided the walls after the pumping step. The background of velocity changing collisions in our experiments, discriminated from the main signal by its shape, is an indication that these gas collisions may restrict the applicability of our technique. However, an extra temporal selection may be added between the pump and probe [41], as the atoms of interest are not "slow", but truly at a "thermal" velocity (within the possible deviations to a standard M-B law) on their direct flight from the pump to the probe region. One may also minimize the effect of collision on the wall between the pump region and the probe spot by designing the cell with a large void between the pump region and the probe spot. This would limit only the exploration of small R values (*i.e.* $R \leq R_{void}$), and hence of relatively fast normal velocities.

At last, aside from its fundamental aspects on the effect of atom-surface interaction onto the gas behavior in the Knudsen layer, further studies following our work should also provide information on the effective possibilities to get truly narrow lines in sub-Doppler techniques close to an interface (thin cell spectroscopy [22, 38], selective reflection spectroscopy [22]), while the state-of-the art of these techniques remains limited to at least several hundredths of the Doppler width. Also, extensions to cells bearing a window with a nanostructured surface would obviously be of interest, to study in clean conditions the effect of the desorption geometry on the detailed velocity distribution.

**ACKNOWLEDGEMENTS**

This work, experimentally supported by BNSF grant (Bulgaria) # MU 02/17, was initiated after calculations by late S. Saltiel. We thank D. Sarkisyan and his group (Ashtarak, Armenia) for the cell preparation.

Fig. 1: Geometry of a spatially-separated pump-probe experiment applied on a 3-level atomic system (schematically shown), illustrating the cell thickness (L) and the internal radius R of the pump. The probe size is assumed to be negligible as in the theory —see figs 2 and 3— while in experiments, R tends to be limited to the free zone between the central probe and the pump ring. The pump and the probe beams can propagate in the same or opposite directions. The choice of a normal incidence, notably for the probe, helps to discriminate the pumped atoms reaching the probe spot after a complex trajectory (scattering, collisions, ...)

Fig. 2: Calculated distribution of the normal atomic velocities $v^\perp$ for atoms detected in the probe region, after a free flight from the pump region, assuming R/L = 10 (left) or R/L = 30 (right), and a thermal velocity $u_{th}$ = 200 m/s. Note that the velocity selection imposed by the cell geometry is softer than the rough estimate $v^\perp < u_{th}$ (R/L)$^{-1}$, obtained assuming that $u_{th}$ is exactly the one-dimensional atomic velocity, and not only the most probable one in a random distribution. The dashed (red and black) vertical line assumes a truncation, with atoms below $|v^\perp| < 4$ m/s excluded from the distribution, *e.g.* because of the surface physics: the truncation has more dramatic effects for R/L = 30 than for R/L = 10. The data are the same as originally published in ref. 42.

Fig. 3: Predicted variations of the amplitude of the probe absorption, as a function of the pump-probe relative distance R/L, after normalization by R (to compensate for the expected decrease in R$^{-1}$). Predictions shown for a M-B-distribution, and for truncated M-B

distributions ($|v^\perp| < v_{\text{excl}}$, as indicated for the excluded velocities $v_{\text{excl}}$ and for $u_{\text{th}} = 200$ m/s). The data are the same as originally published in ref. 42.

Fig. 4: Picture of the 60 μm thick -internal thickness- cell (produced by the D. Sarkisyan group, Ashtarak, Armenia). The macroscopic size of the cell is ~2 x 3 cm. The Cs reservoir surrounded by a metal piece and a thermocouple is also visible.

Fig. 5: Scheme of the experimental set-up.

Fig. 6: Radial analysis of the intensity of the pump ring, by moving an apertured detector ($\varnothing = 0.2$ mm) transversally to the pump cone.

Fig. 7: Amplitude of the (pump-induced) modulated probe absorption at the peak on the 4-4 component as a function of the pump power. Measurements performed for the indicated pump-probe spatial separations and a weak probe (~ 1 μW, *i.e.* ~ 5 μW/mm²). The linear growth at low pump power (dashed lines to guide the eye) is replaced by a plateau above 1.5 mW, independently of the pump-probe separation.

Fig. 8: Frequency spectrum (increasing frequency from left to right) of the (pump-induced) demodulated probe transmission for various pump-probe separation (as indicated). The hyperfine components are indicated on the top of the figure (left: F=3 → F' ={2, 3, 4} and right: F=4 → F' ={3, 4, 5}. The respective separation for the hyperfine components is 150, 200 MHz, and 200, 250 MHz. The vertical scale is the level of the pump-induced

transmission. The experiments are performed for an overall pump power ~1.5 mW, and a probe power (1μW) weak enough to ensure linearity.

Fig.9: Amplitude of the hyperfine peaks of fig.8 counted after subtraction of the broad background as a function of the pump probe separation R for the F= 3→ F' = {2,3,4} manifold (on top) and for the F= 4→F'= {3,4,5} one (on bottom). The pump intensity is 1.5 mW, the probe intensity is weak (1 μW)  and ensures linearity of the probe absorption.

Fig. 10: Same as fig.9, presented with normalized amplitudes as a function of the pump probe separation R (for comparison with fig. 3). The normalization includes the R/L factor, and the average value —including the R/L factor— of each hyperfine component has been normalized to unity of purposes of comparison (see text).

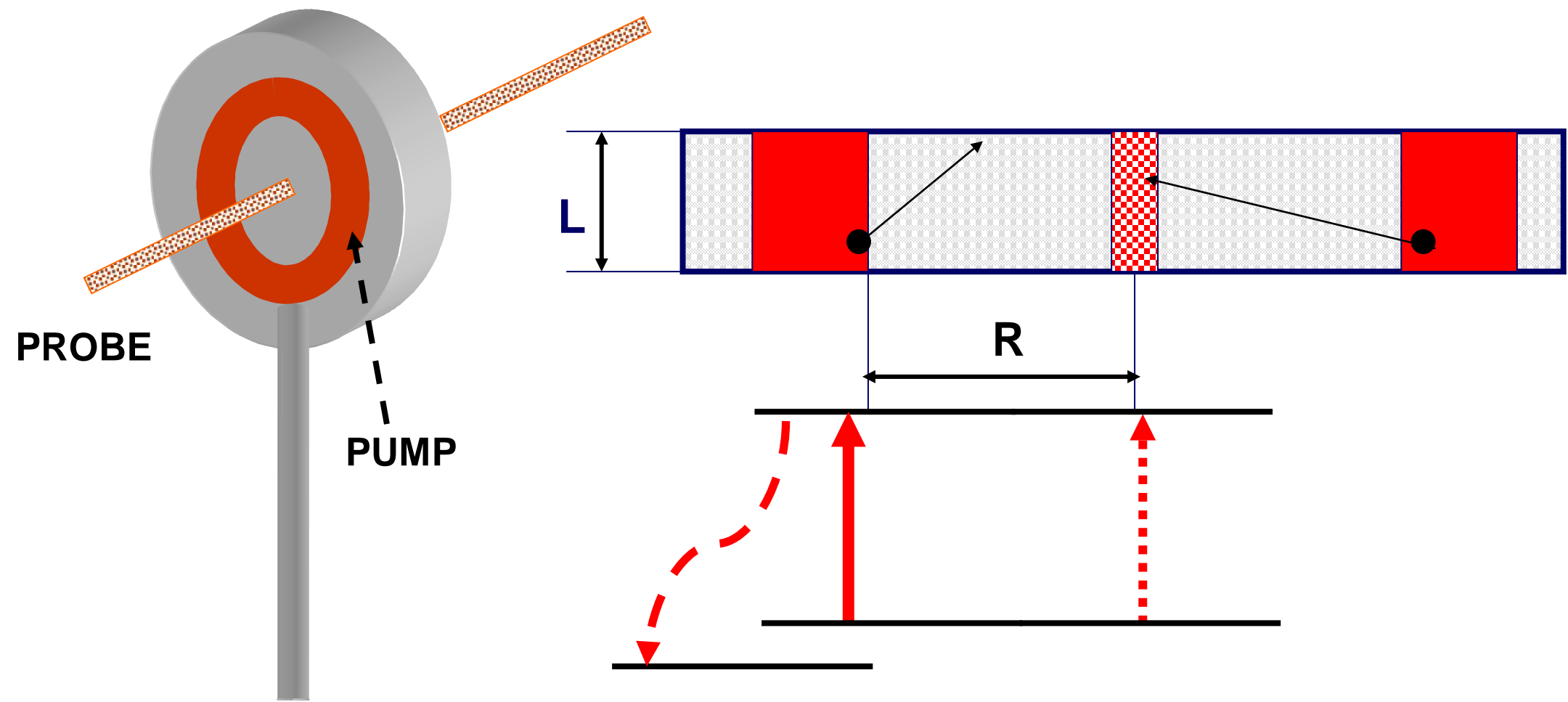

PROBE

PUMP

L

R

Figure 1

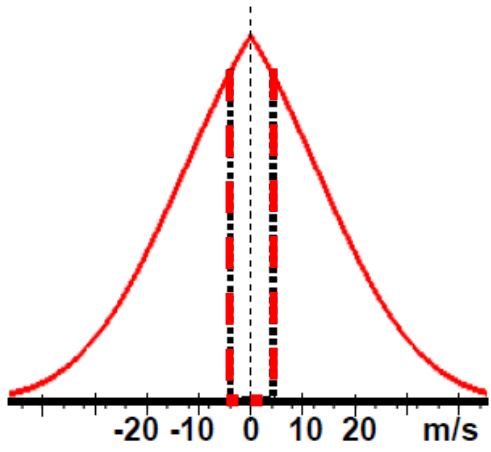 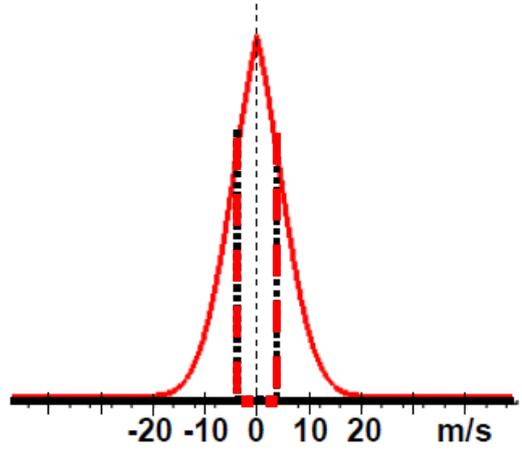

Figure 2

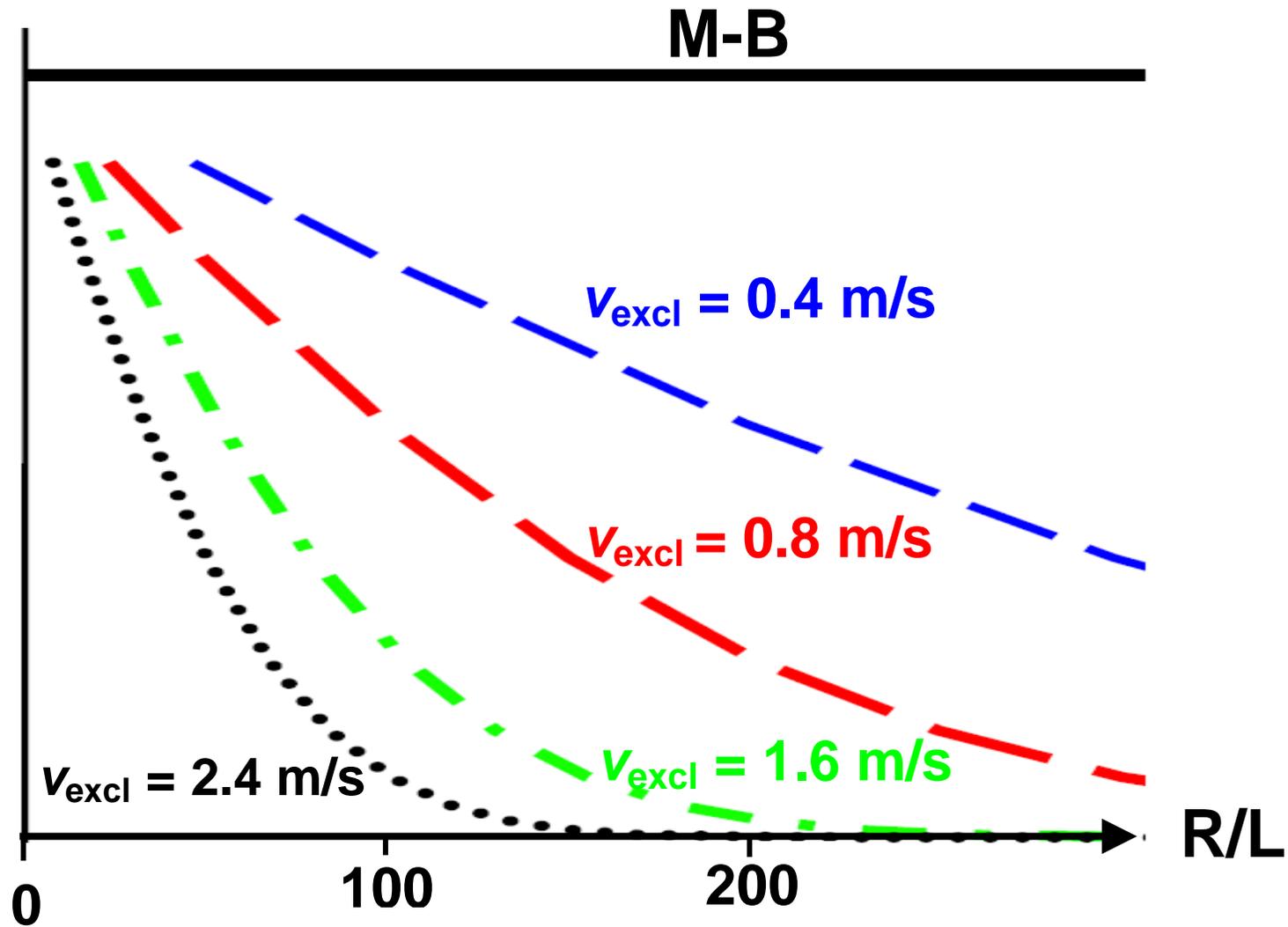

Figure 3

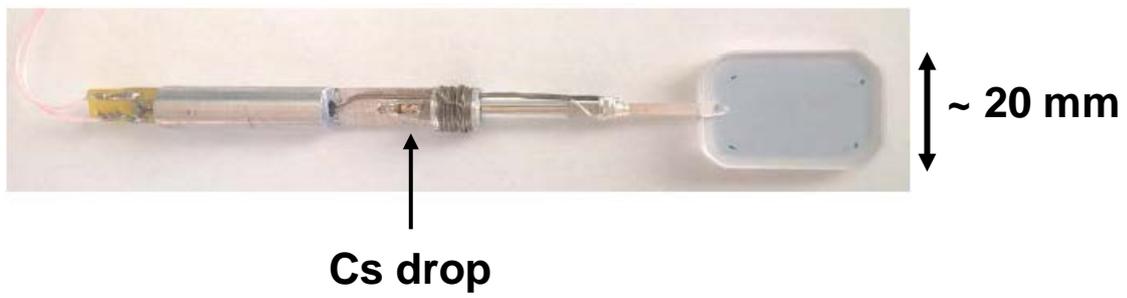

**Cs drop**

~ 20 mm

Figure 4

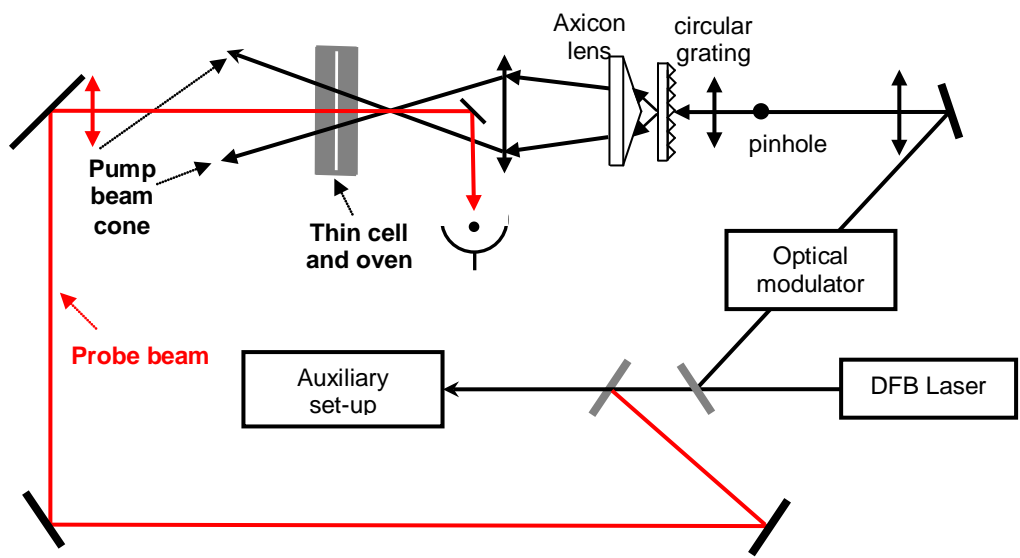

Figure 5

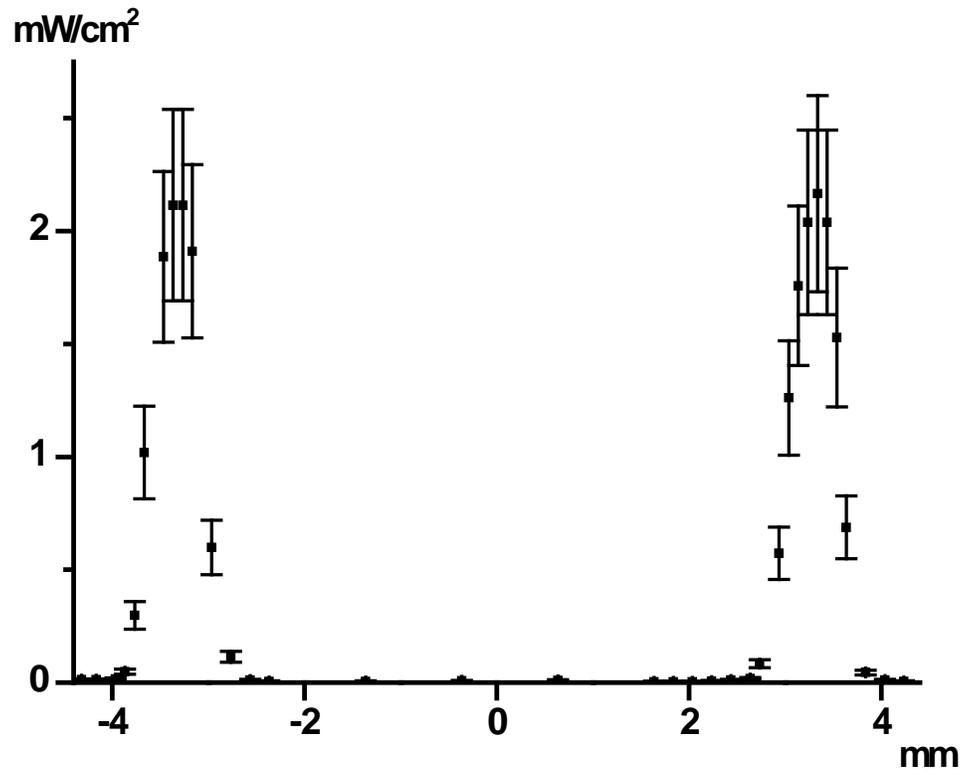

Figure 6

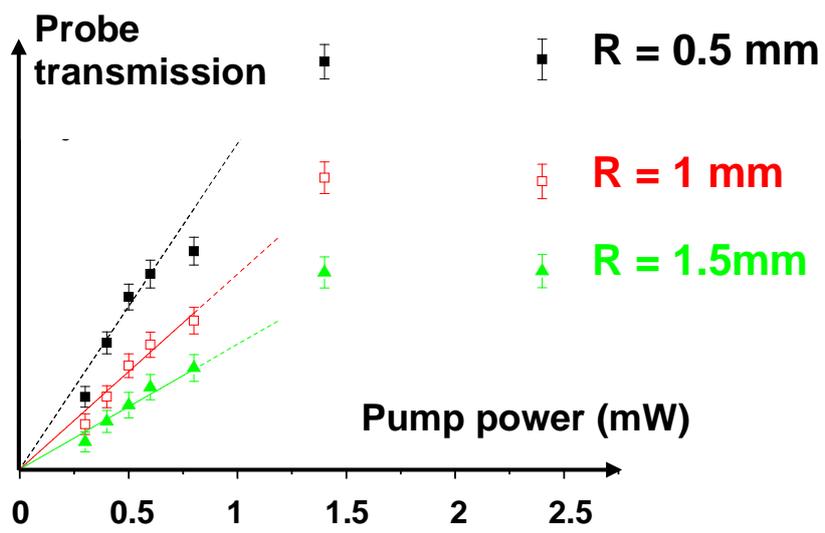

Figure 7

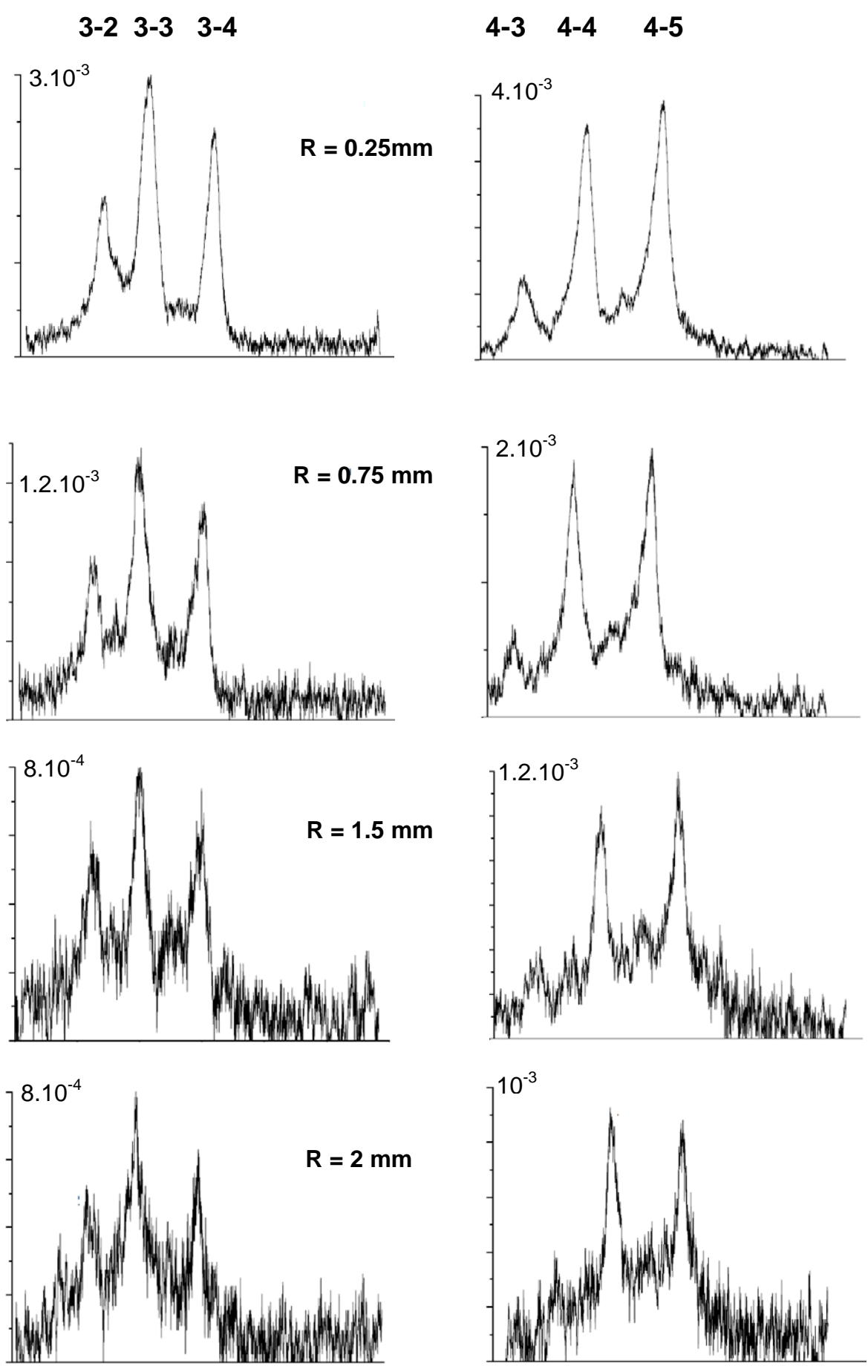

Figure 8

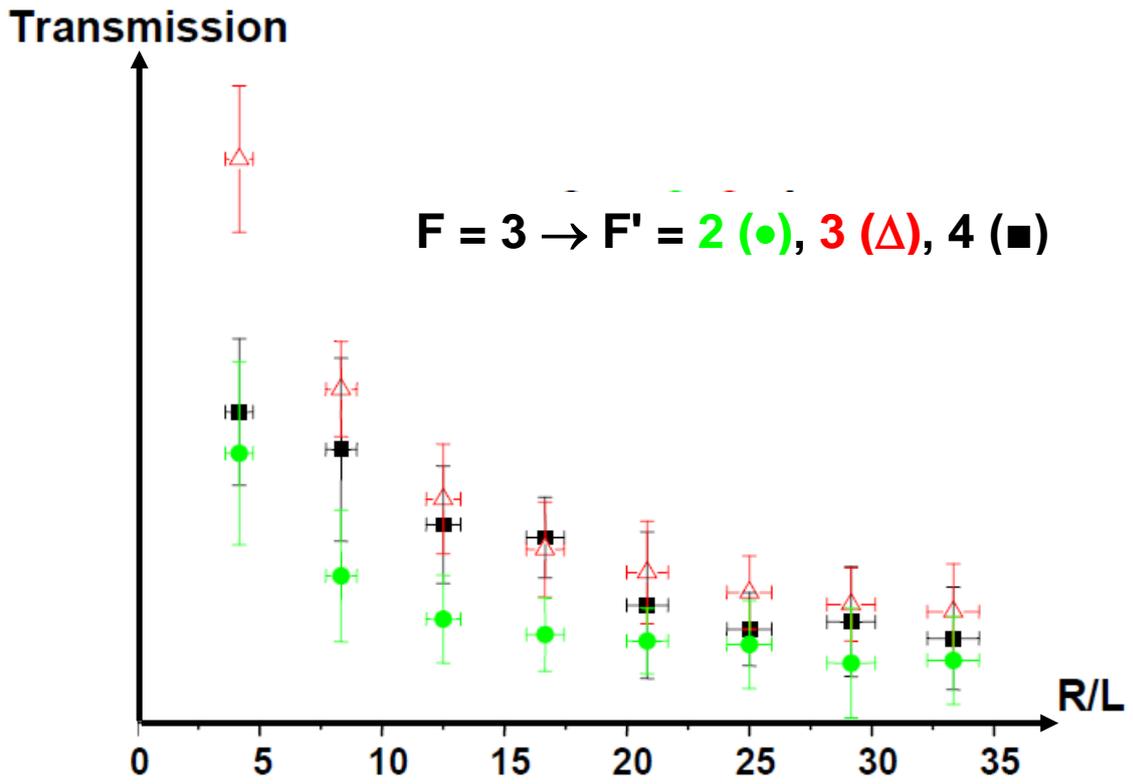

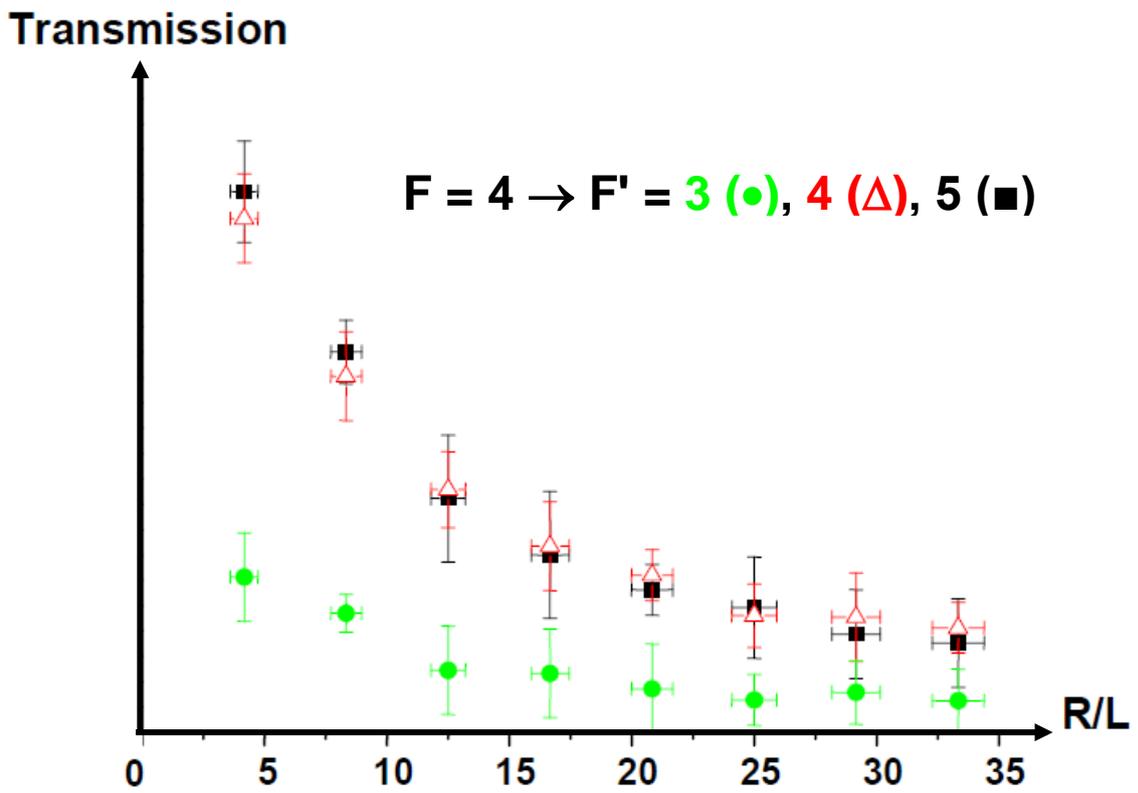

Figure 9

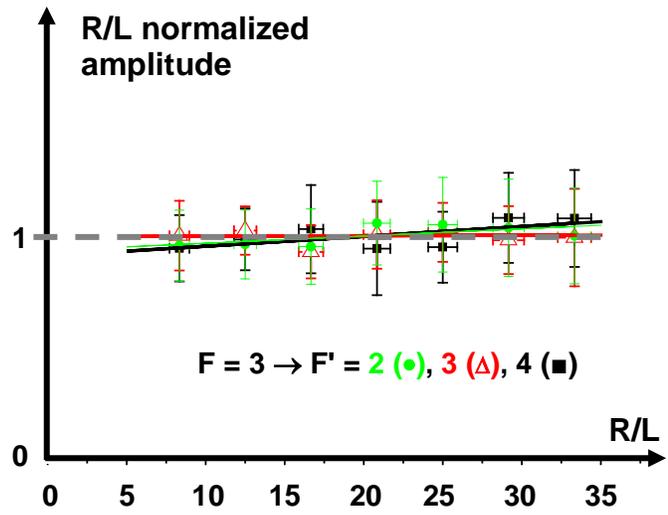

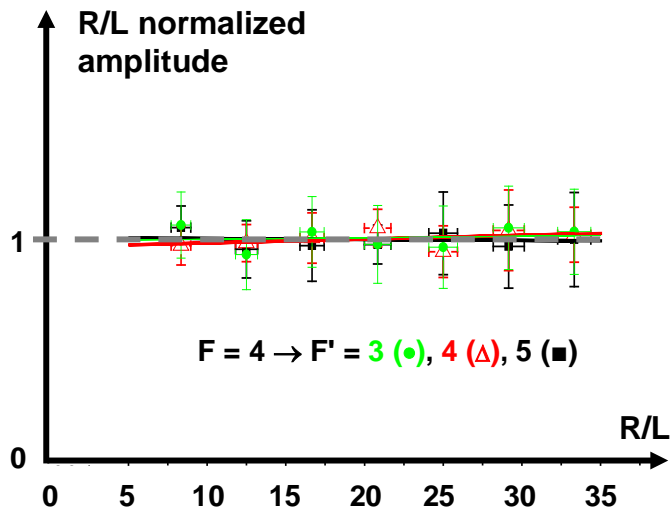

Figure 10